\documentclass[iop,revtex4]{emulateapj}
\slugcomment{{\sc Accepted to ApJ:} September 10, 2013}
\usepackage{lscape}

\shorttitle{A search for pulsations in short GRBs}
\shortauthors{Dichiara et al.}

\begin{document}


\title{A search for pulsations in short gamma--ray bursts to constrain
their progenitors}


\author{S. Dichiara\altaffilmark{1}, C. Guidorzi\altaffilmark{1},
F.~Frontera\altaffilmark{1,2}, L.~Amati\altaffilmark{2}}
\altaffiltext{1}{Department of Physics and Earth Sciences, University of Ferrara,
   via Saragat 1, I-44122, Ferrara, Italy}
\altaffiltext{2}{INAF--IASF Bologna, via Gobetti 101, I-40129, Bologna, Italy}







\begin{abstract}
We searched for periodic and quasiperiodic signal in the prompt emission
of a sample of 44 bright short gamma--ray bursts detected with {\em Fermi}/GBM,
{\em Swift}/BAT, and {\em CGRO}/BATSE. The aim was to look for the observational signature
of quasiperiodic jet precession which is expected from black hole--neutron star
mergers, but not from double neutron star systems. Thus, this kind of search holds the
key to identify the progenitor systems of short GRBs and, in the wait for
gravitational wave detection, represents the only direct way to constrain the
progenitors.
We tailored our search to the nature of the expected signal by properly stretching
the observed light curves by an increasing factor with time,
after calibrating the technique on synthetic curves.
In none of the GRBs of our sample we found evidence for periodic or quasiperiodic signals.
In particular, for the 7 unambiguously short GRBs with best S/N we obtained
significant upper limits to the amplitude of the possible oscillations.
This result suggests that BH--NS systems do not dominate the population of short GRB
progenitors as described by the kinematic model of \citet{Stone13}.
\end{abstract}


\section{Introduction}
Several lines of evidence suggest that short duration gamma--ray bursts
(hereafter, SGRBs; durations $T_{90}\lesssim2$--$3$~s), or at least a sizable fraction
of them, have a cosmological origin and are the electromagnetic counterpart
to the coalescence of compact binary systems, such as double neutron stars (NS)
or neutron star and black hole (BH; e.g., see \citealt{Nakar07,Berger11}
for reviews; see also \citealt{Fong13}; \citealt{Berger13}; \citealt{Tanvir13}).
During the merging, an accretion disk is thought to be produced by the tidal
disruption of a NS around a more compact NS or before a NS is swallowed by a BH.
Either way, eventually the system evolves towards the formation of
a BH with a debris torus around it. The resulting neutrino--cooled accretion
flow leads the hyperaccreting BH to develop a collimated outflow into a pair
of anti--parallel jets (e.g., see \citealt{LRR07}).

A potential means to distinguish between NS--NS and NS--BH mergers
concerns the signature of the disk and jet precession in the electromagnetic signal,
i.e. the SGRB itself. In the case of a NS--BH merger, precession is expected
for a tilted disk and jet due to Lense--Thirring torques from the BH spin
(\citealt{Stone13} and references therein).
These authors (hereafter, SLB13) assumed thick disks precessing as solid
body rotators and built upon numerical relativity simulations of this
kind of mixed mergers. According to their results, for a reasonable set of
values in the parameter space, i.e. BH spin and mass, disk viscosity,
misalignment angle between the accretion disk and the BH equatorial plane,
a quasiperiodic modulation in the $\gamma$--ray signal is to be expected
for a sizable fraction of NS--BH mergers.
The predicted precession period $T_{\rm p}$ increases with time proportionally
to $t^{4/3}$ due to viscous spreading of the disk and, for a given mixed
compact system, starts from a few tens ms
at the beginning of the SGRB, and ends with about one order of magnitude
longer values. The average expected number of cycles is just a few, typically
$N_{\rm cycles}\lesssim 10$. In all scenarios they considered, these two
observables lie in the range $4.5\lesssim\langle N_{\rm cycles}\rangle\lesssim7.5$
and $30$~ms$\lesssim\langle T_{\rm p}(t_{1/2})\rangle\lesssim100$~ms, where
$T_{\rm p}(t_{1/2})$ is the half--way precession period for a given merger.

The aim of this letter is to search for this kind of quasiperiodic signal
in the data of the brightest SGRBs detected with the {\em Fermi}
Gamma--ray Burst Monitor (GBM; \citealt{Meegan09}), the {\em Swift}
Burst Alert Telescope (BAT; \citealt{Barthelmy05a}), and the
{\em Compton Gamma Ray Observatory} Burst And Transient Source Experiment
(BATSE; \citealt{Paciesas99}), exploiting the exquisite time resolution
available with these instruments.
This search offers the only direct way to observationally
distinguish between the two classes of progenitors based on their
electromagnetic emission and naturally complements the forthcoming
gravitational wave studies.
The paper is organized as follows: data selection is described
in Section~\ref{sec:data}. The technique we set up to carry out a dedicated
search is outlined in Section~\ref{sec:proc}. Results and their discussion
follow in Sections~\ref{sec:res} and \ref{sec:disc}, respectively.

\section{Data selection}
\label{sec:data}

\subsection{Sample selection} 
\label{selection}

We took all the events observed by the {\em Fermi}/GBM from July 2008 to December 2012.
For each GRB we extracted and summed the 1--ms light curves of the two most illuminated NaI detectors
in the 8--1000~keV energy band with the {\sc heasoft} package (v6.12) following
the {\em Fermi} team threads.\footnote{http://fermi.gsfc.nasa.gov/ssc/data/analysis/scitools/gbm\_grb\_analysis.html}
Light curves affected by spikes due to the interactions of high--energy
particles with the spacecraft were rejected \citep{Meegan09}.
We derived the $T_{90}$ and  $T_{5\sigma}$ time intervals, where the boundaries of the
latter correspond to the first and the last bin whose counts exceed the 5$\sigma$ signal
threshold above background.

We selected the SGRBs by requiring $T_{90}<3$~s
\footnote{The usual boundary value of $T_{90}=2$~s, which was inherited from the BATSE catalog,
must not be taken too strictly, the two populations of short and long being partially overlapped.
Moreover, this value strongly depends on the detector passband and triggering criteria, as
proven by {\em Swift}/BAT, which detected several SGRBs with $T_{90}>2$~s (e.g. \citealt{Barthelmy05b}).}
, and ended up with 160 GRBs, 18 out of which
having a minimum signal--to--noise (S/N) ratio of 20, as computed over the $T_{5\sigma}$ interval.
As far as the $T_{90}$ distribution is concerned, our selected sample of S/N$>20$ SGRBs is
representative of the full sample of SGRBs, as suggested by a Kolmogorov--Smirnov test.

The same selection criteria were applied to the {\em Swift}/BAT sample using all the events
detected up to early June 2013.  We found 30 GRBs with $T_{90}<3$~s, 12 out of which passed the
final S/N$>20$ threshold. The mask--weighted light curves had previously been extracted
from the event files following the BAT team threads and concern the 15--150~keV detector passband. 
In addition to the 1--ms light curves, for the two brightest events of the sample, namely
051221A and 120323A, we used $0.1$~ms resolution, to explore the very high--frequency
behavior.

From an initial sample of 61 BATSE SGRBs with high S/N we excluded all the cases for which the
time--tagged event (TTE) data did not cover the entire profile. Unfortunately, several bright
bursts were excluded, because the onboard memory could record only up to $32,768$ events around
the trigger time. Consequently, we were left with 14 SGRBs whose profiles were extracted in the
20--2000~keV energy range.

Summing up, our final sample includes 44 (18 {\em Fermi}, 12 {\em Swift}, and 14 {\em CGRO})
SGRBs with high S/N ($>20$). A finer subdivision of the final sample is provided in
the following section, aimed at establishing how genuinely short each selected burst is.

\subsection{Short vs. intermediate GRBs} 
\label{s:classes}
Evidence for the existence of a third group of GRBs with intermediate durations and hardness
ratios between short and long ones was found by several authors for different data sets
(e.g., \citealt{Horvath98,Mukherjee98,Horvath08,Huja09,Ripa09,Horvath09}; but see also \citealt{Koen12}).
In this context, we adopted the classification procedures obtained by \citet{Horvath06} for {\em CGRO}/BATSE
and by \citet{Horvath10} for {\em Swift}/BAT
to assess the nature of our selected sample of bursts, based on the combination of hardness ratio (HR)
and $T_{90}$. We assigned each GRB a probability $p$ of belonging to the short group through the
``indicator function'', out of the three classes: short, intermediate, and long.
As expected, all GRBs had negligible probability of belonging to the long group. 
We defined as ``truly SGRB'' (T--SGRB) the GRBs with $p>0.9$. The GRBs with $0.8<p<0.9$
are defined as ``likely SGRB'' (L--SGRB), whereas the remaining cases ($p<0.8$) were conservatively
classified as ``possibly intermediate'' (P--IGRB). Actually, several members of the P--IGRB group are more
likely to be genuine short than intermediate bursts.
However, our choice was aimed at assuring the least possible contamination with ambiguous cases.
%
\begin{figure}
\includegraphics[width=9cm]{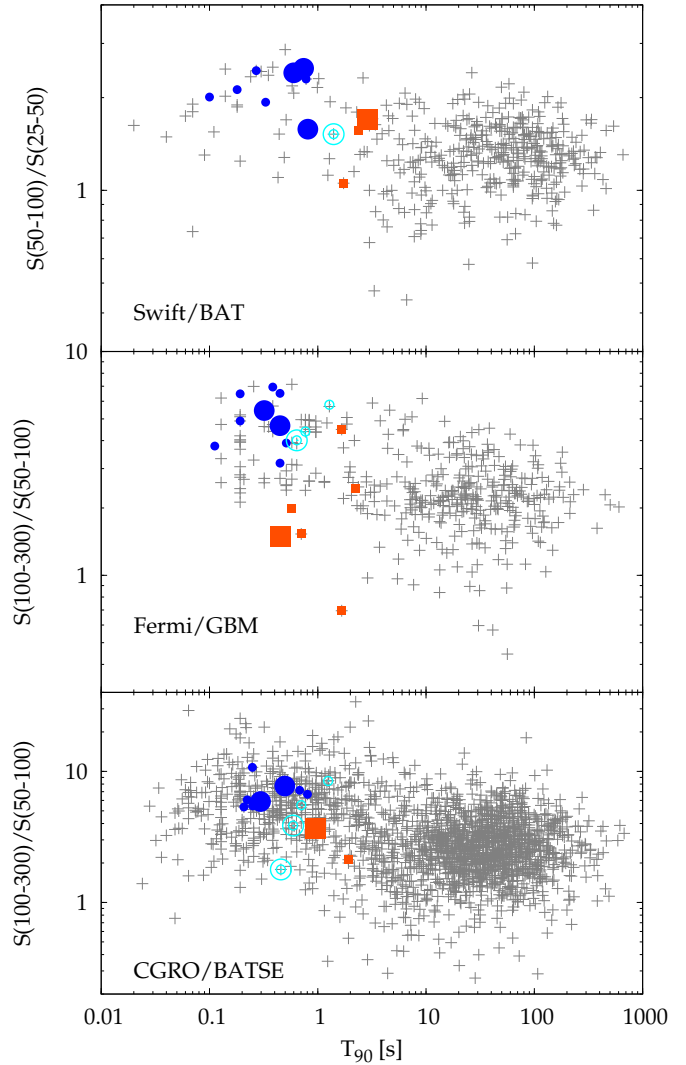}
\caption{ HR--$T_{90}$ diagram for the three data sets: {\em Swift}/BAT ({\em top}),
{\em Fermi}/GBM ({\em mid}), {\em CGRO}/BATSE ({\em bottom}). Each panel shows other catalog GRBs
(crosses) for comparison. Filled circles, empty circles, and squares correspond to T--SGRBs,
L--SGRBs, and P--IGRBs, respectively. Big (small) symbol sizes refer to whether each GRB can
(cannot) provide useful constraints on the possible presence of pulsations using the stretched PDS technique
(Section~\ref{sec:proc}).}
\label{f:classes}
\end{figure}
Figure~\ref{f:classes} shows the HR--$T_{90}$ diagram for the three different data sets: each panel
compares the properties of our selected GRBs with those of the corresponding catalog: \citet{Sakamoto11}
for {\em Swift}/BAT, \citet{Paciesas12} and \citet{Goldstein12} for {\em Fermi}/GBM, and
\citet{Paciesas99} for {\em CGRO}/BATSE.
The HR values for the {\em Swift}/BAT sample were calculated as the fluence ratio in the bands
(50--100~keV)/(25--50~keV) as in \citet{Sakamoto11}, while (300--100~keV)/(50--100~keV) was adopted for the
{\em Fermi}/GBM, and {\em CGRO}/BATSE sets.
To compute the membership probability for the GRBs detected with the {\em Fermi}/GBM, we used the
same parameters used for {\em CGRO}/BATSE owing to the similar energy passbands. Although in principle
this may lead to some misclassified {\em Fermi}/GBM GRBs, in practice the two {\em Fermi} T--SGRBs 
appear to be robustly so (big filled circles in the mid panel of Fig.~\ref{f:classes}).

\section{Data analysis procedure}
\label{sec:proc}
We studied the power density spectrum (PDS) of each light curve in two different ways.
PDS were calculated adopting the Leahy normalization \citep{Leahy83}.
To fit the PDS we used the technique set up by \citet{Vaughan10} based on a Bayesian 
treatment with Markov Chain Monte--Carlo techniques.
Two analytical models were assumed to describe the PDS continuum: a simple power--law plus
constant (hereafter, {\sc PL}),
\begin{equation}
S_{\rm PL}(f) = N\,f^{-\alpha} + B\;,
\label{eq:mod_pl}
\end{equation}
or a broken power--law plus constant (hereafter, {\sc BPL}),
\begin{equation}
S_{\rm BPL}(f) = N\,\Big[1 + \Big(\frac{f}{f_{\rm b}}\Big)^{\alpha}\Big]^{-1} + B
\qquad ,
\label{eq:mod_bpl}
\end{equation}
whose low--frequency index is fixed to zero. In either model the constant term
accounts for the uncorrelated statistical (white) noise. A likelihood ratio test
is used to establish the best model for each PDS.
This technique is particularly suitable to the temporal signal of SGRBs, because it
searches for (quasi)periodic features superposed to a red--noise process and, as such,
can confidently estimate both the best fit parameters of the PDS continuum and
the significance of possible features superposed to it, taking into account the
uncertainties of the model in a self--coherent way.
Moreover, the thresholds for possible periodic features correspond to $2$ and $3\sigma$
(Gaussian) probabilities of a statistical fluctuation and already account for the
multi--trial search over the whole range of explored frequencies in each individual PDS.
It is worth noting that power approximately fluctuates around the model according
to a $\chi^2$ distribution with 2 degrees of freedom, i.e. more wildly than a Gaussian
variate. Actually, the true distribution deviates from a pure $\chi^2$ in that the
model itself is affected by uncertainties. This is properly taken into account by
the procedure in determining the threshold for a given significance
(see \citealt{Vaughan10} and references therein for further details).

\begin{figure}
\includegraphics[width=9cm]{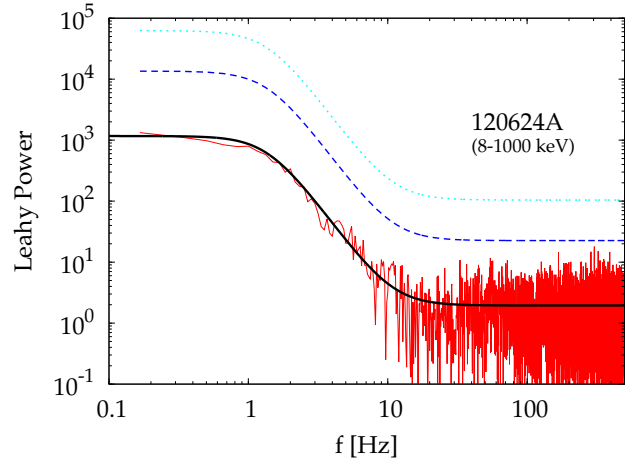}
\caption{The PDS of GRB\,120624A fitted using a Bayesian approach.
The black solid line represents the best fit model ({\sc bpl} in this case) while
dashed and dotted lines give the probability thresholds at 4.5\% and 0.27\% to find a
statistical fluctuation higher than these levels over the whole PDS, respectively.
Confidence levels account for the multi--trial frequencies searched within a given PDS.}
\label{f:2}
\end{figure}
The first search was performed on the observed light curves with uniform binning of
1~ms as they were observed. Hereafter, times are referred to the detector trigger time.
We carried out the same analysis in two different time intervals:
i) from $-3$ to $+3$~s; ii) over the $T_{5\sigma}$ interval.
For the BATSE sample the analysis was carried out just over the $T_{5\sigma}$ intervals
due to the limited memory of TTE data.
The two choices correspond to a fixed temporal range (and, therefore, equal frequency resolution)
and to a S/N--driven scheme, respectively.
For three GRBs, namely 110705A, 120323A, and 130603B, we chose a time interval of 2~s,
spanning from $-1$ to $+1$~s instead of the $T_{5\sigma}$, to properly model the continuum shape.
For 051221A and 120323A we manually selected the time intervals where the analysis was
carried out to exploit at the full the $0.1$--ms time resolution available in these cases:
from $-0.80$ to $+1.20$~s, and from $-0.01$ to $+0.87$~s, respectively. 
These intervals were chosen to optimize the search for possible signals.
Hereafter, we refer to this search as the canonical one, since it does not modify the
light curves so as to account for the increasing precession period expected by SLB13.
Figure~\ref{f:2} shows an example of PDS with the best fit model.
Analogous searches which were already performed in the kHz frequency range in
previous data sets of SGRBs, provided only upper limits to the amplitude
of possible pulsations \citep{Kruger02}.
In the absence of any positive detection of periodic signal, we derived the ${2\sigma}$
upper limits to the amplitude of detectable periodic pulsations for the frequency range
of 10--30~Hz. We expressed this value in terms of fractional amplitude by normalizing
the amplitude limit to the peak count rate of each GRB.

We also performed a second, more sensitive search on the PDS of the same light curves after a
proper stretching of the time axis. To this aim, we devised a technique which was tailored
for the expected signal. For each GRB, we took the $T_{5\sigma}$ interval boundaries
and associated two corresponding precession periods: let $t_0$ and $t_1$ the
start and end times of the $T_{5\sigma}$ interval and let $T_{p,0}$ and $T_{p,1}$
the corresponding precession periods, respectively.
We stretched the time axis according to the continuously increasing $T_{p}$ as
described by Eq.~(\ref{eq:Tp})
\begin{equation}
T_{p}(t)\ =\ T_{p,0}\ \Big(1+\frac{t-t_{0}}{t_{\rm s}}\Big)^{4/3}\;,
\label{eq:Tp}
\end{equation}
where the constant $t_{\rm s}$ is defined as
\begin{equation}
t_{\rm s}\ =\ \frac{t_1-t_0}{(T_{p,1}/T_{p,0})^{3/4} - 1}\;.
\label{eq:ts}
\end{equation}
The values of $T_{p,0}$ and $T_{p,1}$ were chosen so as to match the typical values obtained by SLB13
(typically values were $T_{p,0}=0.01$ and $T_{p,1}=0.6$~s).

We calculated the new count rates in each of the new temporal bins starting from the
original photon arrival times at the detector.
Earlier bins at $t<t_0$ were left unaffected. We attributed a fictitious duration of 1~ms
to the new bins.
We made sure the new bins corresponded to a number of 5 bins per precession period.
This automatically implies that a possible quasiperiodic pulsation such as that described
by Eq.~(\ref{eq:Tp}) should correspond to a frequency $5/2=2.5$ times as small as the
Nyquist one (i.e., 200~Hz in our case) in the stretched PDS.

For each SGRB of our data set, we preliminarily carried out the same analysis on a
set of synthetic curves which were derived from a smoothed version of the original
light curve of the SGRB. The smoothed version was then modulated with different
values of fractional amplitude with a periodic signal with a period varying according
to Eq.~(\ref{eq:Tp}). For each SGRB we determined the minimum amplitude
for which the PDS of the synthetic stretched light curve gave a 2-$\sigma$ detection.
We also searched the synthetic PDS adopting slightly different trial $T_{p,0}$ and
$T_{p,1}$ from the exact values used to build the corresponding stretched curves.
As a result, the detection did not crucially depend on the choice of trial $T_{p,0}$ and
$T_{p,1}$ within a given range. This check is important since this is the case for
real curves for which the possible true periods are unknown a priori.
Further details on how synthetic light curves were generated and on the calibration
of this technique are given in Appendix~\ref{sec:app}.
Hereafter, we refer to this search as the stretched PDS one.

\section{Results}
\label{sec:res}
The canonical search identified just a couple of SGRBs with power exceeding the $2\sigma$
threshold (Gaussian units) in one frequency bin each. The chance probability of a
$2\sigma$ fluctuation occurring within a given PDS is $4.5$\%.
Out of 44 different PDS, the expected number of $>2\sigma$ fluctuations is $1.98$,
i.e. in agreement with the observed number of two cases. Hence, no evidence for the presence
of periodic or quasiperiodic signal was found.
\begin{figure}
\includegraphics[width=8.5cm]{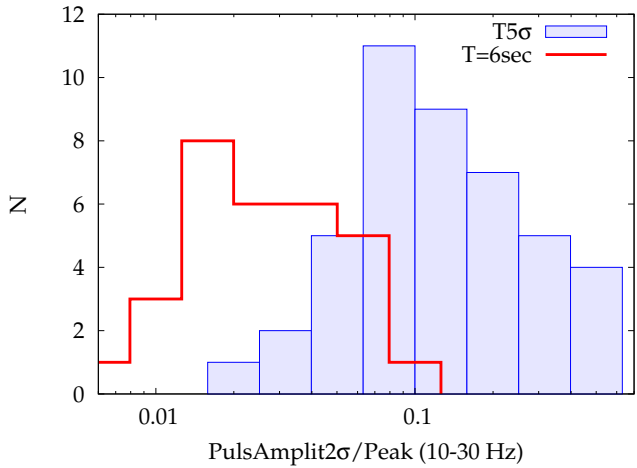}
\caption{Distribution of the minimum detectable pulsation amplitude normalized to peak in
the canonical PDS search. Two cases are shown: fixed time (solid) and $5\sigma$ (shaded) intervals. 
They refer to the 10--30~Hz frequency range.}
\label{f:upplim}
\end{figure}
In the absence of detection, for each GRB we derived a $2\sigma$ upper limit to the fractional
amplitude averaged out over the frequency range of interest, i.e. from 10 to 30~Hz.
The amplitude is normalized to the peak count rate of each SGRB.
The average minimum detectable amplitude depends on the time interval the PDS is calculated:
it clusters around a 3\% (17\%) of the peak for the fixed ($5\sigma$) time interval
(Fig.~\ref{f:upplim}).

\begin{figure}
\includegraphics[width=8.5cm]{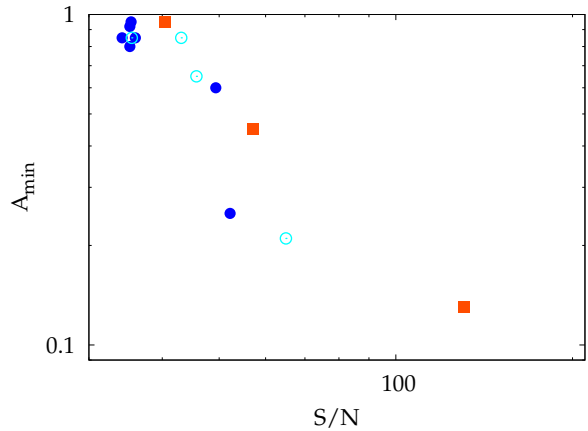}
\caption{Minimum detectable fractional amplitude for an increasing precession period
for 14 SGRBs, as determined from simulations in the stretched PDS search. Same symbols as
in Figure~\ref{f:classes} are used.}
\label{f:upplim_vs_SN}
\end{figure}

Likewise, we did not find any evidence for the quasiperiodic signals in the stretched PDS
search. However, as the calibration on synthetic curves has shown, we could obtain useful
upper limits to the pulsational amplitude for the four, five, and five SGRBs with highest
S/N detected by {\em Fermi}, {\em Swift}, and {\em CGRO}, respectively.
This reduced sensitivity with respect to the canonical search is a consequence of the
low number of expected cycles coupled with the statistical quality of the data.
Figure~\ref{f:upplim_vs_SN} displays the $2\sigma$ upper limits to the fractional amplitude
for a modulation with an increasing precession period superposed to the overall profile
of each SGRB as in Eq.~(\ref{eq:Tp}) as a function of S/N for these 14 events.
With reference to the short/intermediate classification provided in Section~\ref{s:classes},
7 out of these 14 GRBs are T--SGRBs, while the remaining 4 and 3 are L--SGRBs and
P--IGRBs, respectively.
As shown in Figure~\ref{f:upplim_vs_SN}, even neglecting the P--IGRB group our results do
not change in essence, although the reduced number of events demands caution in generalizing
them to larger samples of GRBs.
The burst with the highest S/N and most stringent upper limit to the fractional
amplitude corresponds to GRB\,120323A detected with {\em Fermi}/GBM and it is a
P--IGRB, so the probability of being a misclassified intermediate GRB is not negligible.
Still, it is worth noting that its probability of being a genuine SGRB is 78\% against a
mere 22\% of being intermediate.

\begin{figure}
\includegraphics[width=9cm]{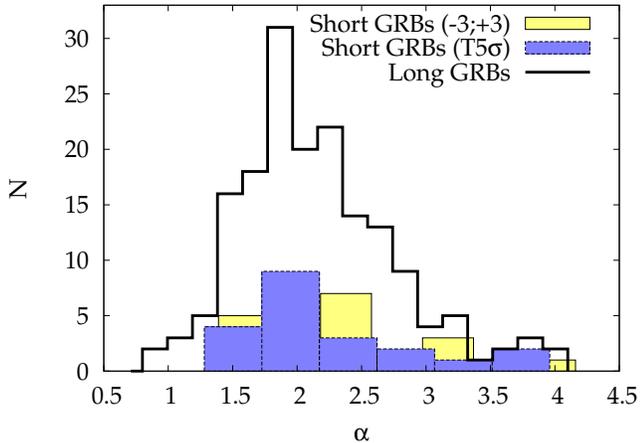}
\caption{Distribution of the PDS slope as derived from the $[-3;3]$~s interval (light shaded),
and the $5\sigma$ interval (dark shaded). Also shown is the same distribution for a sample of
170 long GRBs (solid line; Dichiara et al., in prep.).}
\label{f:alpha}
\end{figure}

Although the QPO search has given negative results, an interesting product of the canonical
search is the continuum properties for an ensemble of bright SGRBs, which is studied here for the
first time. Figure~\ref{f:alpha} shows the distribution of the power--law indices for both
{\sc pl} and the {\sc bpl} models, upon selection of the most accurately measured
values ($|\sigma(\alpha)|<0.5$).
A comparison with analogous results obtained on a sample of long {\em Fermi}/GBM GRBs (Dichiara
et al. in prep.) shows no outstanding difference in the power--law index distribution between short
and long GRBs. Yet the small number of SGRBs lacks in sensitivity to reveal fine differences.

For the SGRBs whose PDS is best fit with a broken power--law, the break frequency
is mostly connected to the overall duration of the main spike, whose timescale is predominant
in the total PDS of SGRBs.

\section{Discussion and Conclusions}
\label{sec:disc}
The canonical search for periodic or quasiperiodic signal did not yield any detection,
in agreement with previous analogous searches \citep{Kruger02}, down to a limiting peak--normalized
amplitude which is typically around 10--20\%  when the PDS is calculated over the $5\sigma$ time
interval.

In addition, we devised and calibrated a technique to detect the signature
of a periodic signal potentially hidden within the time profiles of some SGRBs, characterized
by a continuously increasing period, from a few tens ms up to a fraction of a second or so throughout
the duration of SGRB.
This kind of signal has theoretically been predicted in the case of a mixed merger (NS--BH),
where the tilted jet and accretion disk with respect to the BH spin is expected to
cause the jet precession and a periodic gamma--ray signal in the prompt emission such as that
described above (SLB13).
Likewise, no significant detection at $2\sigma$ out of a sample of 44 SGRBs was obtained by
our tailored technique, named the stretched PDS search, either.
However, we could extract useful upper limits to the
fractional amplitude of such a modulated signal for 14 GRBs, with values distributed
from 10 to 90\%. When we exclude the 3 GRBs which appear to have a non--negligible
($p>0.2$) probability of belonging to the intermediate duration group, the results do not
change in essence.
The reduced sensitivity of the stretched PDS search compared with that of
the canonical one is due to smaller numbers of expected cycles, which couple with a more
critical dependence on S/N, as revealed by the synthetic curves used for calibration.

An interesting outcome of our canonical PDS search concerns the continuum properties of the
PDS for an ensemble of bright SGRBs (see Table in electronic format).
Unlike the case for long GRBs (e.g., see \citealt{Dichiara13} and references therein), this
is the first time we could usefully study these properties for SGRBs, whose study has been
hampered so far by lower S/N with respect to long GRBs.
This was also made possible by the Bayesian procedure
that was recently proposed by \citet{Vaughan10} to properly model the PDS of time series
affected by a strong red noise component, such as the case of SGRBs' time histories (e.g.,
see \citealt{Huppenkothen13}).
Two alternative models were adopted: a simple or a broken power--law in addition to the white
noise constant. A preliminary comparison with the analogous properties of a sample
of bright long GRBs (Dichiara et al. in prep.) reveals no striking difference between the
two power--law index distributions (Fig.~\ref{f:alpha}). Regardless of the PDS continuum
interpretation, this may suggest a common general mechanism which rules the shock formation
and the gamma--ray emission production.

The implications of our results do not allow us to rule out the physical scenario envisaged
by SLB13 as the possible interpretation of the prompt emission of SGRBs for two main reasons.
First of all, the sample of SGRBs for which our non--detection is meaningful is still statistically
too small to draw firm conclusions. This is even more so when one neglects the few GRBs
which could belong to the intermediate duration group.
Secondly, the possibility that the few cases of interest
could correspond to either other kind of mergers, such as NS--NS, or mixed mergers with
unfavorable space parameters, such as the accretion disk viscosity or the misalignment
angle between jet axis and BH spin, is not negligible for just a few cases.
Furthermore, according to the recent physical classification proposed by \citet{Bromberg13},
there could be collapsar events disguised as SGRBs, whose presence could partially explain
the observed lack of evidence for the pulsations expected for NS--BH mergers.
Nonetheless, in addition to being the first attempt of a dedicated search on a valuable
data set, our analysis indicates that such mixed systems might not be a dominant
fraction among the population of currently detected SGRBs, at least as envisaged in
the model by SLB13.
A definitive answer will come from a larger sample with comparable statistical quality
in combination with the wealth of information that will be independently gathered through
the study of gravitation wave radiation.

\appendix
\section{Calibration of the stretched PDS search}
\label{sec:app}
For each SGRBs we carried out a series of simulations aimed at calibrating the sensitivity of
our stretched PDS search. We first binned the original curve to a rough resolution so as to
reduce the high--frequency variability (both real and statistical fluctuations). The smoothed
version of the light curve was then obtained by interpolation of the coarse binned curve by
means of C--splines.
To simulate the predicted periodicity we modulated a smoothed version of the original light curve
with a sinusoidal signal assuming the temporal evolution of $T_{p}$ of Eq.~(\ref{eq:Tp}).
Specifically, to obtain the synthetic light curves we preliminarily had to calculate the
pulsational phase as a function of time, $\phi(t)$. Since $T_p$ continuously varies with time,
we had to integrate the infinitesimal relation $d\phi=2\pi dN = 2\pi dt/T_{p}$, where $dN$ is
the infinitesimal increment to the total number of cycles starting from $t_0$.
Using Equation~(\ref{eq:Tp}) one obtains
\begin{equation}
\phi(t)\ =\ 2\pi\ \int_{t_0}^{t}\ \frac{dt'}{T_p(t')}\ =\ \frac{6\pi t_{\rm s}}{T_{p,0}}\ \Big[1 - \Big(1+\frac{t - t_{0}}{t_{\rm s}}\Big)^{-1/3}\Big]\;.
\label{eq:phi}
\end{equation}
Equivalently, the number of cycles at time $t$, $N(t)$ is given by
\begin{equation}
N(t)\ =\ \frac{\phi(t)}{2\pi}\ =\ \frac{3\,t_{\rm s}}{T_{p,0}}\ \Big[1 - \Big(1+\frac{t - t_{0}}{t_{\rm s}}\Big)^{-1/3}\Big]\;.
\label{eq:N}
\end{equation}
The final number of cycles is given by Eq.~(\ref{eq:N}) at $t=t_1$ and can be conveniently expressed
as 
\begin{equation}
N\ =\ \frac{(t_1-t_0)}{T_{p,0}}\ \frac{3\, x^3}{1 + x + x^2}\;,
\label{eq:Nfin}
\end{equation}
where we defined $x=(T_{p,0}/T_{p,1})^{1/4}$.
The trivial case of constant periodicity ($T_{p,1}=T_{p,0}$) is easily recovered, being
$N=(t_{1}-t_{0})/T_{p,0}$.
Finally, statistical noise was added to the synthetic light curves, which were then processed
exactly in the same as real curves according to the stretched PDS search described in Section~\ref{sec:proc}.

\acknowledgments
This work was supported by PRIN MIUR project on ``Gamma Ray Bursts: from progenitors to physics
of the prompt emission process'', P.~I. F. Frontera (Prot. 2009 ERC3HT).
The useful comments of the anonymous referee are gratefully acknowledged.
We thank Raffaella Margutti for reading the manuscript and for useful comments.

\pagestyle{empty}
\clearpage
\begin{turnpage}
\begin{deluxetable}{llrcccccccccccr}
\tabletypesize{\scriptsize}
\setlength{\tabcolsep}{0.01in} 
\tablecolumns{15}
\tablecaption{Best-fitting model and parameters for each SGRB of the total sample.\label{tbl-1}}
\tablewidth{0pt}
\tablehead{\colhead{GRB} & \colhead{Model} & \colhead{$\log{N}$} & \colhead{$\log{f_{\rm b}}$} & \colhead{$\alpha$} & \colhead{$B$} & \colhead{$p(T_R)$\tablenotemark{a}} & \colhead{$p_{\rm AD}$\tablenotemark{b}}  & \colhead{$p_{\rm KS}$\tablenotemark{c}} & \colhead{$PulseA_{2\sigma}/Peak$} & \colhead{$T_{90}$} & \colhead{$t^{\rm start}$} & \colhead{$t^{\rm stop}$} & \colhead{HR\tablenotemark{g}} & \colhead{$p(Short)$} \\
    &       &           & (Hz)             &          &     &                      &                          &                        &                        &                        &                        &                        &                        &}

\startdata
$051221A$\tablenotemark{d} & {\sc bpl}  & $ 4.487 _{ -1.278 }^{+ 7.780 }$ & $ -0.732 _{ -4.961 }^{+ 0.838 }$ & $ 1.717 _{ -0.110 }^{+ 0.119 }$ & $ 1.812 _{ -0.076 }^{+ 0.072 }$ & $ 0.138 $ & $ 0.654 $ & $ 0.884 $ & $ 0.014 $ & $ 1.370 $ & $ -3.000 $ & $ +3.000 $ & $1.522\pm0.074$ & $0.841$\\
$060313$\tablenotemark{d} & {\sc bpl}  & $ 3.066 _{ -0.350 }^{+ 0.519 }$ & $ 0.089 _{ -0.375 }^{+ 0.276 }$ & $ 1.968 _{ -0.221 }^{+ 0.247 }$ & $ 1.925 _{ -0.074 }^{+ 0.073 }$ & $ 0.428 $ & $ 0.385 $ & $ 0.366 $ & $ 0.019 $ & $ 0.818 $ & $ -3.000 $ & $ +3.000 $ & $2.491\pm0.151$ & $0.999$\\
$061201$\tablenotemark{d} & {\sc pl}  & $ 1.569 _{ -0.179 }^{+ 0.193 }$ & $              -            $ & $ 1.398 _{ -0.222 }^{+ 0.239 }$ & $ 2.025 _{ -0.070 }^{+ 0.071 }$ & $ 0.942 $ & $ 0.997 $ & $ 0.999 $ & $ 0.042 $ & $ 0.827 $ & $ -3.000 $ & $ +3.000 $ &	$2.299\pm0.299$ & $0.999$\\
$080426$\tablenotemark{d} & {\sc pl}  & $ 1.555 _{ -0.196 }^{+ 0.213 }$ & $               -            $ & $ 2.507 _{ -0.512 }^{+ 0.592 }$ & $ 1.954 _{ -0.060 }^{+ 0.060 }$ & $ 0.535 $ & $ 0.845 $ & $ 0.775 $ & $ 0.043 $ & $ 2.019 $ & $ -3.000 $ & $ +3.000 $ &	$1.055\pm0.12$ & $0.123$\\
$081107$\tablenotemark{e} & {\sc pl}  & $ 1.209 _{ -0.216 }^{+ 0.226 }$ & $              -            $ & $ 2.274 _{ -0.455 }^{+ 0.585 }$ & $ 1.958 _{ -0.059 }^{+ 0.060 }$ & $ 0.233 $ & $ 0.877 $ & $ 0.795 $ & $ 0.065 $ & $ 1.792 $ & $ -3.000 $ & $ +3.000 $ &	$0.695\pm0.200$ & $0.006$\\
$081209$\tablenotemark{e} & {\sc bpl}  & $ 1.902 _{ -0.271 }^{+ 0.361 }$ & $ 0.371 _{ -0.323 }^{+ 0.246 }$ & $ 2.478 _{ -0.632 }^{+ 0.829 }$ & $ 1.985 _{ -0.063 }^{+ 0.064 }$ & $ 0.625 $ & $ 0.488 $ & $ 0.664 $ & $ 0.019 $ & $ 0.960 $ & $ -3.000 $ & $ +3.000 $ & $4.897\pm0.657$ & $0.989$\\
$081216$\tablenotemark{e} & {\sc bpl}  & $ 2.342 _{ -0.225 }^{+ 0.271 }$ & $ 0.385 _{ -0.160 }^{+ 0.131 }$ & $ 6.775 _{ -2.928 }^{+ 5.190 }$ & $ 1.981 _{ -0.057 }^{+ 0.059 }$ & $ 0.819 $ & $ 0.888 $ & $ 0.807 $ & $ 0.019 $ & $ 1.152 $ & $ -3.000 $ & $ +3.000 $ & $4.390\pm0.271$ & $0.854$\\
$081223$\tablenotemark{e} & {\sc pl}  & $ 1.622 _{ -0.198 }^{+ 0.216 }$ & $              -            $ & $ 2.368 _{ -0.454 }^{+ 0.535 }$ & $ 1.980 _{ -0.059 }^{+ 0.060 }$ & $ 0.148 $ & $ 0.927 $ & $ 0.627 $ & $ 0.042 $ & $ 1.536 $ & $ -3.000 $ & $ +3.000 $ &	$1.988\pm0.953$ & $0.747$\\
$090108$\tablenotemark{e} & {\sc bpl}  & $ 2.510 _{ -0.300 }^{+ 0.393 }$ & $ 0.130 _{ -0.205 }^{+ 0.158 }$ & $ 4.692 _{ -1.432 }^{+ 2.080 }$ & $ 1.982 _{ -0.061 }^{+ 0.062 }$ & $ 0.797 $ & $ 0.499$ & $ 0.559 $ & $ 0.024 $ & $ 0.768 $ & $ -3.000 $ & $ +3.000 $ & $1.531\pm0.595$ & $0.500$\\
\enddata
\tablecomments{Uncertainties on best-fit parameters are given at $90\%$ confidence. This table is available in its entirety in a machine-readable form in the online journal.}
\tablenotetext{a}{$p(T_R)$ is the significance associated to statistic $T_R$.}
\tablenotetext{b}{$p_{\rm AD}$ is the significance of the Anderson--Darling test.}
\tablenotetext{c}{$p_{\rm KS}$ is the significance of the Kolmogorov--Smirnov test.}
\tablenotetext{d}{Detected by {\em Swift}/BAT}
\tablenotetext{e}{Detected by {\em Fermi}/GBM}
\tablenotetext{f}{Detected by {\em CGRO}/BATSE}
\tablenotetext{g}{Uncertainty on hardness ratio are given at 1 sigma confidence}
\tablenotetext{h}{In this case the time interval of PDS extraction is larger then the $T_{5\sigma}$ interval to fit properly the continuum shape}
\end{deluxetable}
\end{turnpage}
\clearpage

\end{document}